\documentclass[aps,prb,reprint,superscriptaddress]{revtex4-1}
\usepackage{graphicx}
\usepackage{dcolumn}
\usepackage{bm}
\usepackage{graphics}
\usepackage{color}
\usepackage{ulem}
\usepackage{textcomp}
\usepackage{amsmath,amssymb}

\begin{document}


\title{Theoretical Analysis of Optically Selective Imaging in Photoinduced Force Microscopy}

\author{Hidemasa Yamane}
\email{yamane@pe.osakafu-u.ac.jp}
\affiliation{Department of Physics and Electronics, Osaka Prefecture University, 1-1 Gakuen-cho, Naka-ku, Sakai, Osaka, 599-8531, Japan}

\author{Junsuke Yamanishi}
\affiliation{Institute for Molecular Science and The Graduate University for Advanced Studies (Sokendai), 38 Nishigonaka, Myodaiji, Okazaki, Aichi 444-8585, Japan}

\author{Nobuhiko Yokoshi}
\affiliation{Department of Physics and Electronics, Osaka Prefecture University, 1-1 Gakuen-cho, Naka-ku, Sakai, Osaka, 599-8531, Japan}

\author{Yasuhiro Sugawara}
\affiliation{Department of Applied Physics, Osaka University, 2-1 Yamadaoka, Suita, Osaka 565-0871, Japan}

\author{Hajime Ishihara}
\email{ishi@pe.osakafu-u.ac.jp}
\affiliation{Department of Physics and Electronics, Osaka Prefecture University, 1-1 Gakuen-cho, Naka-ku, Sakai, Osaka, 599-8531, Japan}
\affiliation{Department of Materials Engineering Science, Osaka University, 1-3 Machikaneyama-cho, Toyonaka, Osaka 560-8531 Japan}

\date{\today}

\begin{abstract}
We present a theoretical study of the measurements of photoinduced force microscopy (PiFM) for composite molecular systems.
Using the discrete dipole approximation, we calculate the self-consistent response electric field of the entire sample including the PiFM tip, substrate, and composite molecules.
We demonstrate a higher sensitivity for the PiFM measurement on resonant molecules than by the previously obtained tip-sample distance dependency $z^{-4}$ owing of the multifold enhancement of the field between the localized electric field induced at the tip-substrate nanogap and the molecular polarization.
The enhanced localized electric field induced at the tip-substrate nanogap in PiFM allows high-resolution observation of the forbidden optical electronic transition in dimer molecules.
We investigated the wavelength dependence of PiFM for dimer molecules and obtained images at incident light wavelengths corresponding to allowed and forbidden transitions. 
We reveal that these PiFM images drastically change with the frequency-dependent spatial structures of the localized electric field vectors and resolve different types of nanoparticles beyond the resolution for the optically allowed transitions. 
This study demonstrates that PiFM provides multifaceted information based on microscopic interactions between nanomaterials and light.
\end{abstract}

\maketitle

\section{Introduction}
 Localized surface plasmon (LSP) resonance generated in metallic nanostructures induces a strongly enhanced electric field, which has been employed to realize a variety of functions, such as highly sensitive biosensing \cite{liu2003colorimetric}, hot carrier generation \cite{brongersma2015plasmon}, and optical trapping of nanoparticles \cite{pin2018trapping}.
The LSP-related effects are expected to contribute not only to creating next-generation nanotechnologies but also to the understanding of fundamental physics by probing nanoscale optical responses.
One of the peculiar effects in the optical response induced by LSP resonance is the breakdown of the long-wavelength approximation (LWA) at the single molecular level, wherein the optical forbidden transition occurs owing to the spatial variation of the electric field at the nanometer-scale.
For example, the excitation of quadrupole-like polarization of single-walled carbon nanotubes \cite{takase2013selection} and the 2s-3d forbidden transition in a hydrogen atom \cite{kim2018selection} have been reported.
Among various types of electronic transitions in nanostructures, most transitions are optically forbidden, whereas the allowed transitions are particular ones whose associated polarization patterns are nearly structureless.
However, in various types of optical processes and photochemical reactions, the electronic levels that cannot be reached by the one-photon optical transition presumably play significant roles.
Therefore, unveiling and controlling forbidden optical transitions of nanostructures or at the single-molecular level will enhance the degrees of freedom in the design of optical functions of materials.

To understand the optical response of individual nanostructures, it is crucial to elucidate the microscopic interaction between the near-field light and matter. 
Scanning near-field optical microscopy (SNOM)\cite{zenhausern1995scanning,matsuda2003near,kim2006high,novotny2006near}, tip-enhanced Raman spectroscopy (TERS) \cite{hayazawa2000metallized,zrimsek2017single}, and two-photon-induced photoluminescence (TPL) \cite{imura2004plasmon,imura2005near} are powerful tools for investigating near-field distributions beyond the diffraction limit of light and optical processes of individual nanostructures. 
However, these techniques detect the scattered light propagating from the sample into the far field through the interaction between the near-field light and targeted materials and do not observe the near-field light itself. 
Further, the resolution of aperture SNOM and TPL employing SNOM system is limited owing to the collection and the propagation losses in the optical fiber.
Contrarily, photoinduced force microscopy (PiFM) \cite{rajapaksa2010image,rajapaksa2011raman} detects optical interactions at the nanometer scale as local forces in the near-field, which do not suffer from any type of photo-signal attenuation.
Thus, this technique is promising for directly detecting matter near-field interactions at high resolutions. 
Moreover, PiFM has no difficulties regarding the resolution, such as background scattered light or the losses mentioned above.
Theoretical and experimental studies of cantilever dynamics revealed the nanometer scale sensitivity of photoinduced forces to the tip-sample distance owing to the gradient force component \cite{jahng2014gradient}.
Furthermore, it was demonstrated that the combination of mechanical eigenmodes and laser modulation frequency provides images with clearer contrast\cite{jahng2016quantitative}.
Recently, the heterodyne frequency modulation (heterodyne FM) method \cite{yamanishi2017heterodyne,yamanishi2018heterodyne} was used for PiFM measurements in ultrahigh vacuum to remove photothermal oscillations and achieve high-resolution imaging of less than 1 nm \cite{yamanishi2020optical}.
Additionally, PiFM can acquire information on samples through their specific linear and nonlinear optical responses. 
For example, optical force measurements by atomic force microscopy for the detection of nonlinear optical signals have been theoretically studied \cite{saurabh2014communication}. Further, the selective detection of particular chemical species in the sample \cite{chen2019revealing} and the analysis of features of buried components \cite{almajhadi2017contrast} were performed through the resonant optical responses of samples.

Notably, PiFM can potentially acquire information on photo-excitation processes, including forbidden optical transitions. 
As discussed in theoretical studies \cite{iida2006force,iida2007theoretical}, multifaceted information of excited states of nanostructures can be obtained as a spatial map of induced forces in high resolution under an electronic resonance condition by using the degrees of freedom of incident light, such as polarization, frequency, and wave-vector (incident angle). 
Presently, more elaborate studies of such capabilities are desired, assuming modern PiFM techniques.
Thus, the purpose of this study is to elucidate the possible sensitivity, resolution, and information we can acquire when the state-of-the-art PiFM technique \cite{yamanishi2020optical} is applied to observe forbidden optical transitions.
In Ref. \cite{yamanishi2020optical}, we numerically simulated the PiFM measurements of a dumbbell-shaped quantum dot \cite{kameyama2018enhanced}, whose spatial structures of electronic schemes were designed especially to achieve high performance of the photocatalytic function. 
Using the discrete dipole approximation (DDA) method \cite{purcell1973scattering}, we calculated the total response electric field self-consistently with induced polarization and investigated the forces acting on the tip, treating the strongly enhanced LSP in the metallic nanogap region.
Consequently, we successfully reproduced the frequency-dependent three-dimensional force vector map reflecting the functions of the targeted sample and elucidated the mechanism of achieved resolution in the above experiment. 
In this study, we extend our model system to measure the composite molecules and discuss the information obtained for the relevant cases.

\section{Model and Theory}
\begin{figure}
  \includegraphics[width=\linewidth]{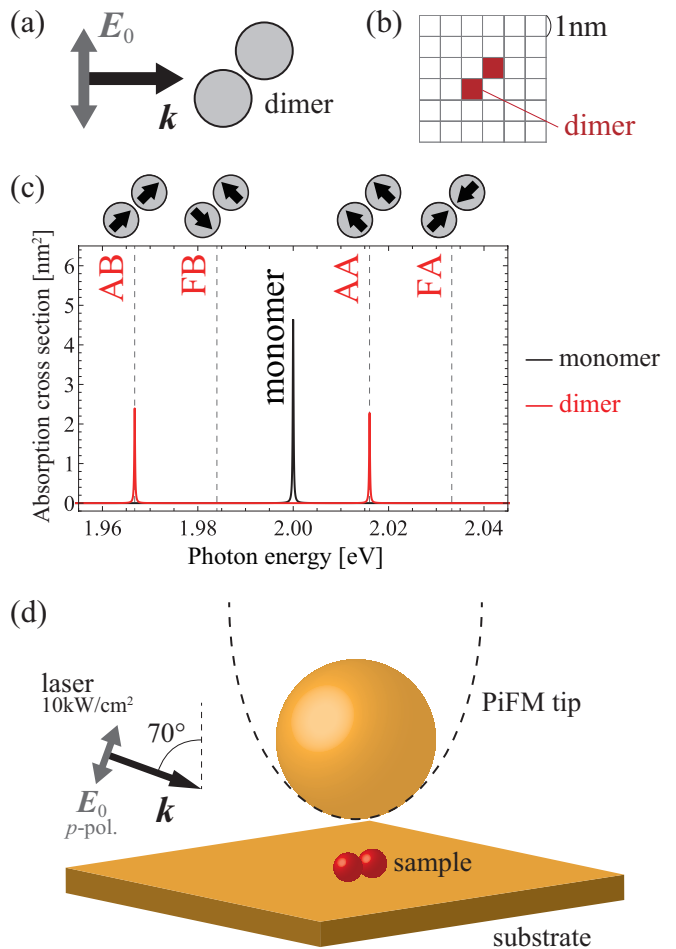}
  \caption{(a) Model geometry of dimer used in calculation. (b) The dimer is represented by two aligned cells, whose susceptibilities are given by the Lorentz model in DDA. (c) Spectra of the induced field intensity at a monomer (black line) and a dimer (red line) in vacuum. The induced field intensity was normalized by the incident field intensity. Vertical lines represent the energies of the allowed bonding (AB), forbidden bonding (FB), allowed antibonding (AA), and forbidden antibonding (FA) states of the dimer. In vacuum, the dimer can be excited at the energies of the AB and AA states under the LWA condition. The circles above the spectra represent dimers, and the arrows in the circles indicate the directions of the dipoles. (d) Schematic of PiFM model. We approximate the metal-coated tip to the gold sphere. \label{fig:model}}
\end{figure}
Here, we assume that the dimer molecule has four kinds of optical transitions according to the type of interaction between its dipoles; these are, in the order from lower to higher energy: allowed bonding (AB), forbidden bonding (FB), allowed antibonding (AA), and forbidden antibonding (FA) states (See Fig. \ref{fig:model}(a)-(c)).
The optical-allowed transitions are associated with the states in which the dipoles on each molecule are in the same phase, and conversely, the optically forbidden transitions are associated with the states where the dipoles are in the opposite phase. 
The latter transitions cannot be induced optically under the LWA condition.
However, when the dimer molecule is in the vicinity of the metal nanogap, the transitions become possible due to a steep electric field gradient \cite{iida2009unconventional}. 
We clarify the incident photon energy dependence of PiFM measurements for dimer molecules and reveal that completely different PiFM images can be obtained for each transition under non-LWA conditions. 
The different images reflect specific spatial structures of the three-dimensional polarization for different transitions enhanced by LSP, induced between the tip and the substrate. 

We calculated the self-consistent total response electric field induced in a nanogap made of a metal tip, a quantum dot, and a metal substrate using DDA. The following equations were used for the calculation:
\begin{align}
\label{eq:E}
\bm{E}(\bm{r}_{i},\omega)&=\bm{E}_{0}(\bm{r}_{i},\omega)+\int_{V}d\bm{r}_{j}\bm{G}(\bm{r}_{i},\bm{r}_{j},\omega)\bm{P}(\bm{r}_{j},\omega),\\
\label{eq:P}
\bm{P}(\bm{r}_{i},\omega)&=\chi({\bm r}_{i},\omega)\bm{E}(\bm{r}_{i},\omega),
\end{align}
where $\bm{E}(\bm{r}_{i},\omega)$ and $\bm{E}_{0}(\bm{r}_{i},\omega)$ represent the total response field and the incident field, respectively; $i$ is the number of cells at the coordinate $\bm{r}_{i}$, and $\omega$ is the angular frequency of the electric field. 
$\bm{G}(\bm{r}_{i},\bm{r}_{j},\omega)$ is the free-space Green function propagating in both the transverse and longitudinal electromagnetic fields. 
$\bm{P}(\bm{r}_{j},\omega)$ is the polarization of the $j$-th cell, and the integral of the second term in Eq. (\ref{eq:E}) represents the field at the $i$-th cell propagated from the polarization at the $j$-th cell, and $V$ is the volume of one cell. 
$\chi({\bm r}_{i},\omega)$ is the optical susceptibility.

We assume that the gold-coated tip is a gold sphere with a diameter of 19 nm, and the gold substrate is a gold thin film with a size of $143\times143\times9$ nm$^{3}$, as depicted in Fig. \ref{fig:model}. 
The metallic structures are assumed to have a Drude-type susceptibility with the parameters of Au,
\begin{align}
\chi_{\rm{Au}}(\omega)=\frac{1}{4\pi}\left[\epsilon_{\rm {Au}}-\epsilon_{0}-\frac{(\hbar\omega_{\rm Au})^2}{(\hbar\omega)^2+i\hbar\omega\left(\gamma_{\rm bulk}+\frac{\hbar V_{\rm F}}{L_{\rm eff}}\right)}\right],
\end{align}
where $\epsilon_{\rm {Au}}$ is the background dielectric constant of the metal, and $\epsilon_{0}$ is the dielectric constant of vacuum. 
$\omega_{\rm Au}$ is the bulk plasma frequency, and $\gamma_{\rm bulk}$ is the electron relaxation constant of bulk gold;
$V_{\rm F}$ is the electron velocity at the Fermi level; 
$L_{\rm eff}$ is the effective mean free path of electrons comparable to the size of the tip diameter.

Regarding the nanoparticles, we assign one cell in DDA to one particle with a size of 1 nm$^3$ that assumes the following Lorentzian-type of susceptibility: 
\begin{align}\label{eq:Lorentz}
\chi_{\rm{np}}(\omega)=\frac{1}{4\pi}\left[\epsilon_{\rm {np}}-\epsilon_{0}-\frac{|\mu^2|/V_{\rm np}}{\hbar\omega_{\rm np}-\hbar\omega-i\gamma_{\rm np}}\right],
\end{align}
where $\epsilon_{\rm {np}}$, $\mu$, $V_{\rm np}$, $\hbar\omega_{\rm np}$, and $\gamma_{\rm np}$ are the background dielectric constant, dipole moment, volume, resonant energy, and damping constant of the nanoparticle, respectively.

The photoinduced force acting on the tip is obtained by the following formula: \cite{iida2008theory}
\begin{align}
\langle {\bm F}(\omega)\rangle=\frac{1}{2}{\rm Re}\left[\int_{V_{\rm t}}d{\bm r}_{i}(\nabla{\bm E}^{*}({\bm r}_{i},\omega))\cdot{\bm P}({\bm r}_{i},\omega)\right],
\end{align}
where the integral is performed over the volume of the tip $V_{\rm t}$.

\section{Sensitivity and Resolution of PiFM}

To reveal the sensitivity of PiFM, we calculated the photoinduced force spectrum and the force curve.
We used the following parameters for the gold structure: $\epsilon_{\rm {Au}}=12.0$, $\epsilon_{0}=1$, $\hbar\omega_{\rm Au}=8.958\ {\rm eV}$, $\gamma_{\rm bulk}=72.3\ {\rm meV}$, $\hbar V_{\rm F}=0.922\ {\rm nm\cdot eV}$, and $L_{\rm eff}=20{\ \rm nm}$ \cite{johnson1972optical}.
We assumed that p-polarized and 10 kW/cm$^2$ plane wave light illuminates the sample at an angle of 70$^\circ$ as the incident laser.
We consider nanoparticles, whose resonant energies are $\hbar\omega_{\rm np}=2.0$ and 2.1 eV, and the dipole moment 10 debye, which is similar to the value for porphyrin-based dye molecules \cite{tsuda2001fully,mukai2001effects}, with the parameters $\epsilon_{\rm {np}}=1.5$, $\gamma_{\rm np}=5$ meV and $V_{\rm np}=1$ nm$^{3}$.
We call them NP1 and NP2, respectively.

In Fig. \ref{fig:spect}(a), the red and green solid lines represent the spectra of the absorption cross-section for the presence of only one nanoparticle NP1 and NP2 in the free space, respectively.
The absorption cross-section was previously defined by \cite{goedecke1988scattering}
\begin{align}
\sigma_{\rm abs}(\omega)=4\pi\frac{\omega}{c}\int_{V_{\rm np}}d{\bm r}_{i}\frac{|{\bm E}({\bm r}_{i},\omega)|^2}{|{\bm E}_{0}({\bm r}_{i},\omega)|^2}{\rm Im}[\chi_{\rm np}({\bm r}_{i},\omega)].
\end{align}

\subsection{Wavelength selectivity of PiFM}
\begin{figure}
  \includegraphics[width=0.8\linewidth]{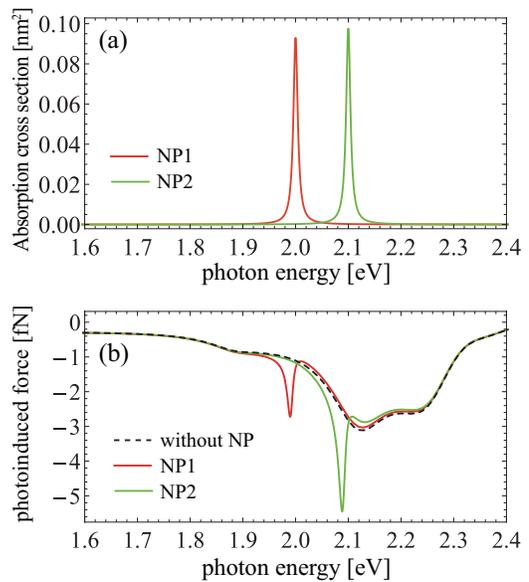}
  \caption{Spectra. (a) Spectra of the absorption cross-section of NP1 (red line) and NP2 (green line) in free space. (b) Photoinduced force spectra for the tip on NP1 (red line), NP2 (green line), and gold film (black line).\label{fig:spect}}
\end{figure}
Fig. \ref{fig:spect}(b) shows the photoinduced force spectra detected by the PiFM tip.
The red and green lines represent the spectra for the presence of only one nanoparticle NP1 and NP2 on the substrate, respectively. 
The center of the tip is just above the nanoparticle and tip-sample distance $d$, defined as the distance between the tip surface and nanoparticle surface, at 1 nm.
The peak positions are red-shifted to 1.990 and 2.088 eV owing to interaction with the LSP.
The black line represents spectrum on the gold substrate, i.e., in the absence of nanoparticles.
The plasmonic resonance of gold yields a broad spectrum.
 
\begin{figure}
  \includegraphics[width=0.8\linewidth]{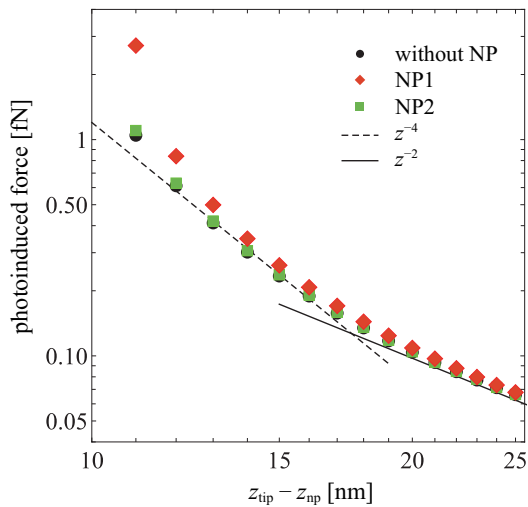}
  \caption{Force curves of PiFM. The attractive force is plotted on a logarithmic graph as positive. Red and green markers represent the force curve on nanoparticles NP1 and NP2, respectively. The black markers represent forces on the gold substrate, which are overlapped by green markers. The photoinduced force is proportional to $z^{-4}$ for short distances and $z^{-2}$ for long distances. 
For NP1, the force curve does not follow the $z^{-4}$ law because of the influence of the resonance and becomes even steeper in the near range.\label{fig:fc}}
\end{figure}

Fig. \ref{fig:fc} shows the force curve of the photoinduced force at laser photon energy $\hbar\omega = 1.990$ eV. 
The horizontal axis depicts the distance between the position of the center of the tip $z_{\rm tip}$ and the nanoparticle $z_{\rm np}$.  
The red, green, and black markers represent the force curve on the NP1, NP2, and gold substrate, respectively.
The diameters of the tip and the nanoparticle were 19 nm and 1 nm, respectively, and they contacted at $z_{\rm tip}-z_{\rm np}=10$ nm ($d=0$ nm).
However, because the cell size of the DDA is set to 1 nm$^3$ cubic, $d$ cannot be below 1 nm in this calculation.
For a long distance, the force curve follows $z^{-2}$ lows owing to the Coulomb force.
The photoinduced gradient force is proportional to $z^{-4}$ by approximating the tip and sample as two point dipoles \cite{jahng2014gradient,ladani2017dyadic}.
In this situation, the laser photon energy is close to the resonant energy of NP1, and the force curve of NP2 is proportional to $z^{-4}$, as in the case of the gold substrate (the absence of NP) for a short distance.
Contrarily, the force curve of NP1 does not follow the $z^{-4}$ law and becomes even steeper in the near range ($z_{\rm tip}-z_{\rm np}<13$ nm) owing to multiple enhancements in the field between the LSP in the nanogap and the polarization of the NP1 resonance.
This indicates that PiFM has high sensitivity for the resonance of target materials.

\subsection{Distinguishing different nanoparticles}
We show that PiFM is advantageous in detecting the presence of nanoparticles, as well as the structure and physical properties of the surface of the sample.
This leads to the identification of defects and impurities within the sample and its surface, as well as the chemical composition and optical response of the sample molecules.

\begin{figure}
  \includegraphics[width=\linewidth]{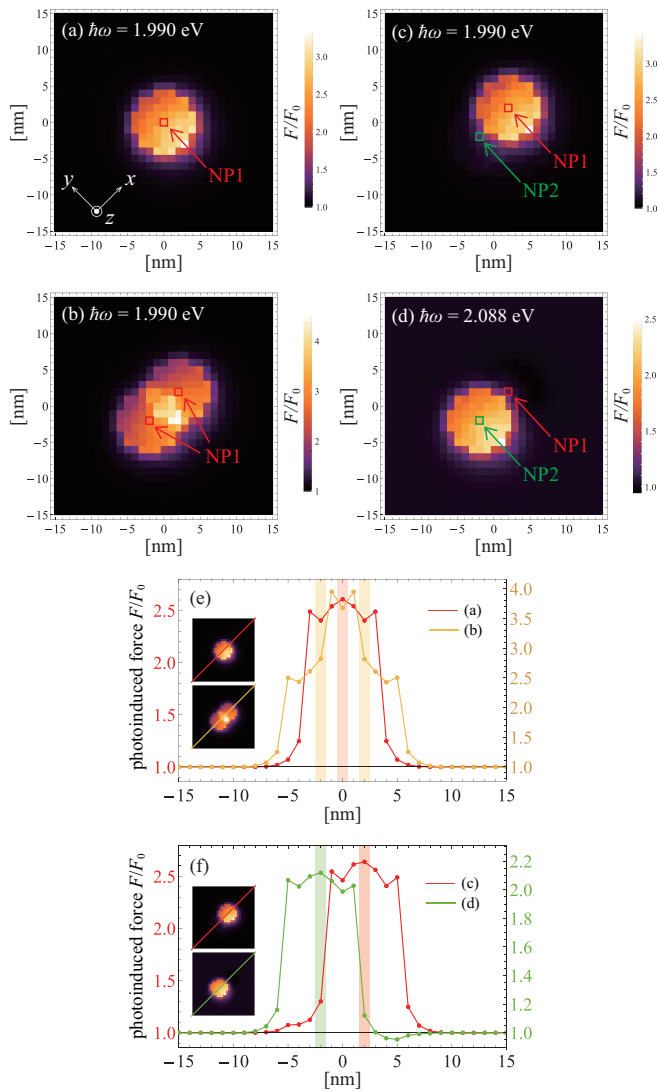}
  \caption{PiFM images. (a) Single nanoparticle PiFM image. The red squares represent the positions of the nanoparticles. (b) Image of two aligned nanoparticles. (c) and (d) show images for different types of nanoparticles with different resonant energies aligned. The incident direction of the laser shifts from positive to negative on the y-axis at an incident angle of 70$^\circ$ and p-polarization. The laser photon energies are 1.990 eV for (a), (b), and (c) and 2.088 eV for (d). (e) depicts the force profiles of (a) and (b), and (f) depicts the force profiles of (c) and (d). The pale-colored bands of red, yellow, and green represent the positions of nanoparticles. \label{fig:PiFM1}}
\end{figure}

We obtained the PiFM image by scanning the tip at a height of 1 nm from the nanoparticles.
Fig. \ref{fig:PiFM1}(a) shows a PiFM image of one nanoparticle (NP1) with laser photon energy $\hbar\omega=1.990$ eV. 
The incident direction of the laser shifts from positive to negative on the $y$-axis at an incident angle of 70$^\circ$ and p-polarization.
The red square in the figure represents the position of the nanoparticle. The color map represents the z-axis component of the photoinduced force acting on the tip.
The force profile on the line in Fig. \ref{fig:PiFM1}(a) is shown by the red line in Fig. \ref{fig:PiFM1}(e).
The PiFM resolution is below 10 nm at full width at half maximum.
This resolution is proportional to the tip diameter.
Fig. \ref{fig:PiFM1}(b) shows a PiFM image in the case where two same kinds of nanoparticles (NP1) are aligned. 
The distance between the two nanoparticles is about 4.24 nm (diagonal length of three cells).
The force profile on the line is shown by the yellow line in Fig. \ref{fig:PiFM1}(e).
As mentioned above, the resolution is approximately 10 nm, such the force profiles overlap, and PiFM cannot differentiate between the two nanoparticles.

However, if two different types of nanoparticles are aligned, we can differentiate between them by PiFM.
Fig. \ref{fig:PiFM1}(c) and (d) show PiFM images of two different kinds of nanoparticles with laser photon energies of $\hbar\omega=1.990$ and 2.088 eV, respectively.
The red square represents nanoparticle NP1, and the green square represents nanoparticle NP2.
In the case of laser photon energy $\hbar\omega=2.088$ eV, NP1 can be measured (Fig. \ref{fig:PiFM1}(c)). 
Contrarily, in the case of the laser photon energy $\hbar\omega=1.990$ eV, NP2 is measured (Fig. \ref{fig:PiFM1}(d)).
Fig. \ref{fig:PiFM1}(f) shows the force profiles of each laser photon energy.
These results imply that PiFM can distinguish different types of nanoparticles beyond the resolution of the tip diameter by tuning the wavelength of the incident laser.

The wavelength selectivity of PiFM can be used to help identify chemical modifications on the surface of protein molecules \cite{dietz2019optical}.
PiFM also has the advantage of being able to observe buried materials \cite{almajhadi2017contrast}, which could play a role, for example, in detecting interlayer impurities in nanoparticle multilayered optical devices, assessing the structure of the underlying layers, and identifying chemical species adsorbed within the porous medium.

\subsection{Damping constant dependence of sensitivity}
\begin{figure}
  \includegraphics[width=\linewidth]{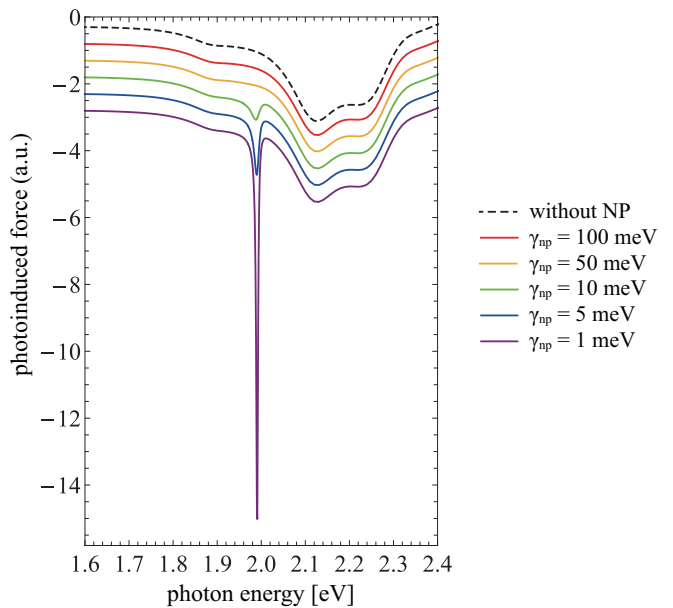}
  \caption{Damping constant of nanoparticle dependence of the photoinduced force spectrum. The dashed black line represents the absence of nanoparticles, and the solid colored lines depict nanoparticles with damping constants $\gamma_{\rm np=}$ 100, 50, 10, 5, and 1 eV. The resonance energy of the nanoparticles is at 2.0 eV, and as the damping constant decreases, the peak appears.\label{fig:damping}}
\end{figure}
We show the damping constant dependence of the photoinduced force spectra in Fig. \ref{fig:damping}.
Their resonance energies are $\hbar\omega_{\rm np}=2.0$ eV, and the damping constants are $\gamma_{\rm np}=$ 1, 5, 10, 50, and 100 meV.
As the decay constant of the nanoparticles decreases, the peaks become larger and narrower.

Even when the damping constant is large, if the two energies are sufficiently far apart such that the peaks of the photoinduced force spectrum do not overlap, they can be observed selectively, albeit with low sensitivity.
Naturally, it is desirable that the line width of the sample is sharp for the PiFM measurement with high sensitivity and high selectivity.
At low temperatures, the scanning tunnel luminescence peaks of the dye molecules have line widths of a few \textmu eV \cite{imada2017single}.
In this case, PiFM allows us to distinguish various molecules and observe various vibrational levels.

\section{Optical Response of Allowed and Forbidden Transitions of a Dimer}
In this section, we analyze the PiFM measurements of a dimer.
We model the dimer as two aligned monomers, whose susceptibilities are described by the Lorentz model Eq. (\ref{eq:Lorentz}) with dipole moment $\mu$ = 10 debye and resonance energy $\hbar\omega_{\rm np}$ = 2.0 eV. Two monomers are coupled with each other via the dipole interaction and form a dimer that can assume four excited transition states: AB, FB, AA, and FA states \cite{iida2009unconventional}. 
For clear energy splitting, we assume the damping constant in susceptibilities to be $\gamma_{\rm np}$ = 0.5 meV.
We show the absorption cross-section of the dimer in free space in Fig. \ref{fig:model} (c).
The bonding and antibonding states of the dimer appear symmetrically split from the resonance state of the monomer.
In the LWA, the states with the same phase of dipoles are allowed and forbidden when the dipoles are in the opposite phase.
We show that completely different PiFM images can be obtained for each transition energy of the incident laser.
These PiFM images reflect the three-dimensional structure of the local electric field vectors.

\subsection{PiFM spectra of dimer}
\begin{figure}
  \includegraphics[width=1.0\linewidth]{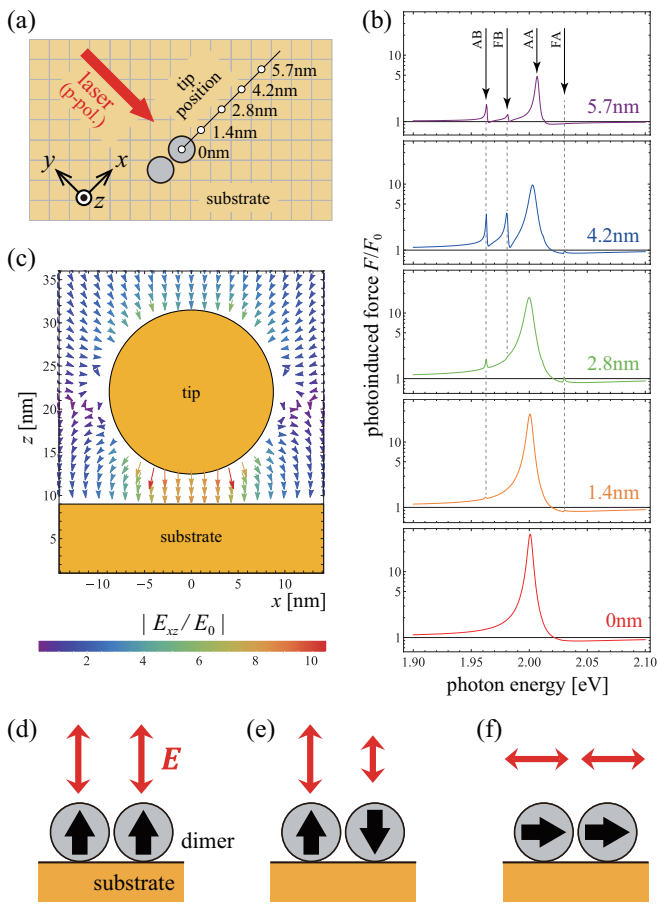}
  \caption{(a) Schematic illustration of dimer on gold substrate and tip position. The center of the tip is represented by white dots. The grid on the substrate represents the cells in DDA. The incident direction of the laser shifts from positive to negative on the $y$-axis at an incident angle of 70$^\circ$ and p-polarization. (b) Photoinduced force spectra. The lines represent the dimer, and each tip position is shown in (a). As the tip moves away from the dimer, the peak corresponding to the AA state becomes smaller, whereas the peaks corresponding to the AB and FB states appear strongly in the range 2.8 - 4.2 nm. (c) Vector map of response field around tip. The color of the vectors represents the magnitude of the response field in the absence of the dimer. (d) When the dimer is just under the tip, the AA state can be excited. (e) When the dimer is away from the tip, the FB state can be excited owing to the gradient of the field. (f) The AB state can be excited by the appearance of the lateral component of the field.
\label{fig:dimer_fspect}}
\end{figure}
We calculated the photoinduced force spectra of the dimer to reveal the PiFM measurement for the allowed and forbidden transitions.
We illustrate the calculation model in Fig. \ref{fig:dimer_fspect}(a).
The dimer is on the gold substrate, and the white dots represent the center of the tip.
The incident direction of the laser shifts from positive to negative on the $y$-axis at an incident angle of 70$^\circ$ and p-polarization.
The tip is just 1 nm above the dimer. Fig. \ref{fig:dimer_fspect}(b) represents the photoinduced force spectra for the different positions of the tip illustrated in Fig. \ref{fig:dimer_fspect}(a).
The colored solid lines represent the dimer.
As the tip moves away from the dimer, the peaks corresponding to the AA state become smaller.
In contrast, the peaks that correspond to the AB and FB states appear around the tip position 2.8 - 4.2 nm; nevertheless, they are observed at 0 nm.
This can be explained by the structure of the electric field response.
Fig. \ref{fig:dimer_fspect}(c) shows the spatial distribution of the response electric field vectors around the tip without the dimer.
Here, we show the case of a 2.0 eV photon energy of the laser. 
Owing to the wide range of plasmon resonance peaks, the vector field hardly changes with a difference of a few tens of meV.
Because the $z$-component of the LSP field is strongly enhanced at the nanogap between the tip and the substrate, the AA state can be excited just under the tip (Fig. \ref{fig:dimer_fspect}(d)).
Contrarily, because of the steep electric field gradient, the FB state can be excited around the tip.
Owing to the shape of the tip, the lateral component of the LSP field appears, which is why the AB state can be excited at 1.4 nm or more.
\begin{figure}
  \includegraphics[width=\linewidth]{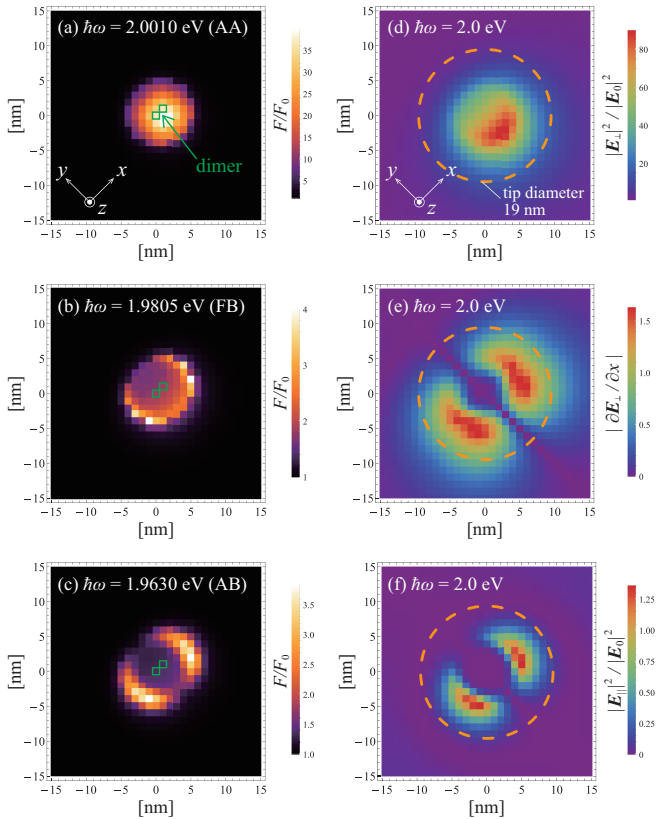}
  \caption{(a), (b), and (c) depict PiFM images of the dimer for AA, FB, and AB states, respectively. The dimer is represented by green squares. (d), (e), and (f) represent the response field distributions under the tip that is absent from the dimer. 
The incident photon energy is $\hbar\omega=2.0$ eV.
In the absence of the dimer, the metallic structure creates a wide range plasmon resonance peaks, which hardly changes with a difference of a few tens of meV.
The yellow dashed line circles represent the tip position and diameter.
(d) is the intensity of the $z$-component of the field, (e) is the $x$-direction gradient of the $z$-component of the field, and (f) is the intensity of the $x$-component of the field.
The incident direction of the laser shifts from positive to negative on the $y$-axis at an incident angle of 70$^\circ$ and p-polarization.\label{fig:PiFM_dimer}}
\end{figure}

\subsection{PiFM images of dimer}
We showed that the excitation selectivity of the dimer depends on the spatial structure of the localized response electric field.
We obtained different PiFM images of the dimer with the incident photon energies corresponding to each transition state.
In this section, we show that the PiFM images reflect the spatial structure of the intensity and polarization of the localized response field.

First, the PiFM image for the AA state is shown in Fig. \ref{fig:PiFM_dimer}(a).
The AA state excitation can be understood in terms of the LWA and depends on the intensity of the electric field in the perpendicular direction with respect to the direction in which the dimers are aligned (i.e., $\bm{E}_{\perp}\equiv(0,E_{y},E_{z})$), as shown in Fig. \ref{fig:dimer_fspect}(d).
The intensity of the electric field $\bm{E}_{\perp}$ distribution on the substrate around the tip is shown in Fig. \ref{fig:PiFM_dimer}(d), where the tip position is represented by the yellow dashed line circle.
The incident photon energy is $\hbar\omega=2.0$ eV; however, the metallic structure without any nanoparticles creates a wide range of plasmon resonance peaks, and hence, it hardly changes with the change of the photon energy by a few tens of meV.
In the PiFM system, the response electric field is remarkably enhanced and localized in the $z$-direction, which is the direction of the gap between the tip and the substrate.
The electric field is slightly distorted owing to the asymmetry caused by the incident direction of the laser ($y$-direction).
This is reflected in the PiFM image (Fig. \ref{fig:PiFM_dimer}(a)).

Subsequently, in Fig. \ref{fig:PiFM_dimer}(b), we show the PiFM image for the FB state.
Considering the dipole structure of the dimer, $\bm{E}_{\perp}$ elements contribute to FB state excitation as well as the AA state, as shown in Fig. \ref{fig:dimer_fspect}(e).
However, this is a plasmonic phenomenon beyond the LWA, and it occurs owing to the steep gradient along the $x$-direction of the electric field $\bm{E}_{\perp}$ on the size scale of the dimer, that is, $\partial \bm{E}_{\perp}/\partial x$, as shown in Fig. \ref{fig:PiFM_dimer}(e).
This effect results in the PiFM image around the dimer, as shown in Fig. \ref{fig:PiFM_dimer}(b).
Because of the size of the dimer itself, the dimer feels a certain electric field gradient, even just below the tip.

Finally, the PiFM image for the AB state is shown in Fig. \ref{fig:PiFM_dimer}(c).
AB state excitation is a phenomenon that occurs in the LWA. However, in contrast to AA and FB, it is contributed by the electric field in the direction of the dimer alignment (i.e., $\bm{E}_{\parallel}\equiv(E_{x},0,0)$), as shown in Fig. \ref{fig:dimer_fspect}(f).
In PiFM, the electric field in the $z$-direction, which is the direction of the gap, is remarkably enhanced, as mentioned above.
Because the bottom of the tip is relatively flat, the horizontal component of the electric field is hardly induced just under the tip.
However, the horizontal electric field is generated away from the nadir of the tip owing to the inclination of its curved surface.
We show the distribution of the intensity of the electric field of $\bm{E}_{\parallel}$ in Fig. \ref{fig:PiFM_dimer}(f).
This distribution is reflected in the PiFM image shown in Fig. \ref{fig:PiFM_dimer}(c).

We assumed a tip diameter of 19 nm and stated that the resolution was approximately 10 nm. 
However, in the PiFM measurement of the dimer, the PiFM images can provide information on localized electric field structures beyond the resolution determined by the tip radius.
As described above, PiFM is a unique optical microscope that not only reveals the geometrical structure and optical property of the sample but also provides the three-dimensional structure of the localized electric field, including spatial distribution and polarization based on the microscopic interaction between the localized electric field and the sample at the nanometer scale.

\section{Summary and Conclusion}
We theoretically analyzed the photoinduced force measurement on nanoparticles by PiFM.
The interaction between the polarization of nanoparticles and the LSP induced at the metal nanogap between the gold tip and gold substrate are self-consistently considered, yielding the total response electric field by the DDA.
We clarified that PiFM exhibits high sensitivity to the resonance energy of nanoparticles.
We obtained the PiFM images of nanoparticles on the gold substrate.
Because of the high sensitivity to the resonance energy, different PiFM images can be obtained by tuning the wavelength of the incident laser. 
Even if different kinds of nanoparticles exist beyond the resolution determined by the tip diameter, PiFM allows the differentiation of these nanoparticles depending on their spectral sharpness. 

In this study, we simulated the PiFM measurement for a dimer molecule.
 Forbidden optical transition excitations are observed by PiFM.
The near-field is induced at the nanogap between the PiFM gold tip and gold substrate.
The three-dimensional structure of the near-field can excite various optical transitions, including transitions forbidden under LWA.
When the dimer is just under the tip, the polarization of the near field is directed to the gap mode and allowed antibonding (AA) transition to be excited.
Around the tip, the steep electric field gradient enables the forbidden bonding (FB) transition excitation.
Further, owing to the tip shape, the lateral component of the polarization of the near field appears, which makes it possible to excite the allowed bonding (AB) transition.
Owing to PiFM sensitivity, the PiFM images corresponding to the respective transitions are acquired completely differently, and they reflect the three-dimensional structure of the near-field PiFM images for AA, FB, and AB corresponding to the vertical component, the lateral gradient of the vertical component, and the lateral component of the near field, respectively.

Unlike other microscopy techniques, such as SNOM \cite{zenhausern1995scanning}, TERS \cite{li2010shell}, and TPL \cite{imura2005near}, PiFM does not require a photo-signal propagating to detectors, but measures the induced force between the dipoles of the sample and probe. 
Therefore, this method does not suffer from any signal attenuation during its propagation. This feature enables a significantly higher resolution compared with other optical microscopes. 
Unlike conventional atomic force microscopes, PiFM can acquire information on electronically excited states through the observation of induced polarization that directly reflects three-dimensional spatial structures of the electric field in the vicinity of the samples. 
A particularly notable point is the possibility of observing optically forbidden transitions. 
If we make complete use of the degrees of freedom of light, including the frequency, polarization, and wavevector, we can obtain multifaceted information on microscopic interaction between nanomaterials and light.
In our previous study \cite{yamanishi2020optical}, we revealed that a picocavity effect \cite{benz2016single} plays an essential role in the high resolution of PiFM, where we used the multi-sized cell DDA method developed in \cite{takase2013selection} to consider the atomic-scale spatial structures on the tip and sample. 
In the present study, we did not use this method to conduct the study with wide parameter regions, avoiding the heavy computational load. 
However, in a following study, we will use the multi-sized cell DDA combined with elaborated calculations of electronic states of molecular systems to demonstrate what we can see by the observation of a single molecule with PiFM, which would reveal the potential capability of PiFM to observe the excitation processes of molecular systems.


\begin{acknowledgments}
This work was supported by JSPS KAKENHI Grant Number JP16H06504 for Scientific Research on Innovative Areas ``Nano-Material Optical-Manipulation''
\end{acknowledgments}

\bibliography{main}

\end{document}